\documentclass[sigconf,authorversion,nonacm]{acmart} %
\AtBeginDocument{%
  }

\usepackage{tikz}
\newcommand{\Cross}{$\mathbin{\tikz [x=1.4ex,y=1.4ex,line width=.2ex, black] \draw (0,0) -- (1,1) (0,1) -- (1,0);}$}%

\usepackage{amsfonts}
\usepackage{mathtools}

\usepackage{multirow}

\usepackage{stfloats}

\usepackage[utf8]{inputenc}
\usepackage[T1]{fontenc}
\usepackage{microtype}

\usepackage{siunitx}

\usepackage{xspace}

\newcommand{\myparagraph}[1]{\smallskip \noindent {\bf #1}.}

\usepackage{tcolorbox}
\newtcolorbox{mybox}[1]{colframe=white!20!black,title=#1,boxsep=1pt,left=5pt,right=5pt,top=5pt,bottom=5pt}

\newcommand{\PSS}{{\it PSS}\xspace}
\newcommand{\PSSO}{{\it $PSS_O$}\xspace}

\begin{document}

\title{Scalable Program Clone Search Through Spectral Analysis}

\author{Tristan Benoit}
\email{benoit.tristan.info@gmail.com}
\affiliation{%
  \institution{Université de Lorraine, CNRS, LORIA}
  \city{Nancy}
  \country{France}
  \postcode{F-54000}
}

\author{Jean-Yves Marion}
\email{jean-yves.marion@loria.fr}
\affiliation{%
  \institution{Université de Lorraine, CNRS, LORIA}
  \city{Nancy}
  \country{France}
  \postcode{F-54000}
}

\author{S\'ebastien Bardin}
\email{sebastien.bardin@cea.fr}
\affiliation{%
  \institution{CEA LIST, Universit\'e Paris-Saclay}
  \city{Saclay}
  \country{France}
}

\begin{abstract}
We consider the problem of program clone search, i.e.~given a target program  and a repository of known programs (all in executable format), the goal is to  find the program in the repository most similar to the target program -- with potential  applications in terms of reverse engineering, program clustering, malware lineage and software theft detection. Recent years have witnessed a blooming in code similarity techniques, yet most of them focus on function-level similarity  and function clone search, while we are interested in program-level similarity and program clone search. %
Actually, our study shows that prior  similarity approaches are either too slow to handle large program repositories, or not precise enough, or yet not robust against slight variations introduced by compilers, source code versions or light obfuscations. 
We propose a novel spectral analysis method for program-level similarity and program clone search called Programs Spectral Similarity (\PSS). 
In a nutshell,  \PSS\ one-time spectral feature extraction is tailored for large repositories, making it a perfect fit for program clone search. 
We have compared the different approaches with extensive benchmarks, showing that  \PSS reaches a sweet spot in terms of precision, speed and robustness. 

\end{abstract}

\begin{CCSXML}
<ccs2012>
   <concept>
       <concept_id>10002978.10002997.10002998</concept_id>
       <concept_desc>Security and privacy~Malware and its mitigation</concept_desc>
       <concept_significance>300</concept_significance>
       </concept>
   <concept>
       <concept_id>10002978.10003022.10003465</concept_id>
       <concept_desc>Security and privacy~Software reverse engineering</concept_desc>
       <concept_significance>500</concept_significance>
       </concept>
 </ccs2012>
\end{CCSXML}
\ccsdesc[500]{Security and privacy~Software reverse engineering}
\ccsdesc[300]{Security and privacy~Malware and its mitigation}

\keywords{binary code analysis, clone search, spectral analysis}

\settopmatter{printfolios=true}
\maketitle

\section{Introduction}

Binary code similarity approaches identify similarities or differences~\cite{SBC} between pieces of assembly code  (e.g., basic blocks, binary functions or whole programs). 
We focus on program-level similarities (coined \textit{program similarity} in the following), that is, computing a similarity index between whole programs which is capable of telling at which degree two programs are similar -- with potential  applications in terms of reverse engineering, program clustering, malware lineage and software theft detection.

\myparagraph{Program clone search} Given a query composed of a \textit{target program} and a repository, the {\it program clone search} ranks repository programs by their program similarity to the target program. The search is successful if the most similar program is a \textit{clone} of the {target} program. These clones may be  (i) compiled with slightly different compiler chains, or (ii) produced from a slightly different version of the source code, or (iii) altered by slight obfuscations. %

\myparagraph{Applications} Searching program clones between x86 or ARM binaries over a large program repository is necessary when the original program written in source code is unavailable, which happens with commercial off-the-shelf (COTS), legacy programs, firmware or malware.
For example, detecting malware clones is a major issue~\cite{MSBB, MAV, SMM, MMDFC}, as   
most malware are actually variants of a few major families active for more than five years\footnote{\url{https://www.cisa.gov/uscert/ncas/alerts/aa22-216a}}.
Another application is the identification of  libraries~\cite{AndroidLibrary,AndroidBloom,LibDX,LibDB, BAT, OSSPolice}, which is both a software engineering issue and a cybersecurity issue due to  vulnerabilities inside dynamically linked libraries.
The problem of library identification, while in between programs and functions in terms of size, is much closer to the case of program clones by its nature, as libraries are not arbitrary collections of functions and require inter-procedural analysis. 
The situation is similar for  patch and firmware analysis~\cite{SPAIN}, or software theft detection~\cite{BAT, OSSPolice, EXPOSE}, which 
also need  to consider a global view of the code. 

\textit{In all these cases, we see function clone search as only a proxy to a problem that is by nature at the level of programs. }

\myparagraph{Prior work} Given its potential applications and challenges, the field of similarity detection  has been extremely active over the last two   decades, starting from the pioneering work of Dullien in 2004~\cite{SCEO,GBCEO} on call-graph isomorphisms  and the popular BinDiff tool for  recognizing similar binary functions among two related executables.  %
Other approaches include for example symbolic methods~\cite{BINHUNT}, 
graph edit distances~\cite{LSMI,ICGC} and matching techniques~\cite{MAV,SMM}. 
Interestingly, the last five years have seen a strong trend toward machine learning based approaches to binary function similarity~\cite{Asm2Vec,AlphaDiff,CODEE,GEMINI,SAFE}.  
\textit{Overall, most prior work focuses on function clone search and function-level similarity.}

\myparagraph{The challenges}   
Program clone search presents specific challenges compared to standard function similarity. 
(1) As already stated, it requires comparing programs, i.e.~much larger objects than functions, hence similarity checks must be scalable in typical program sizes;  
(2) We do not consider two programs taken in isolation, but a target program and a (possibly large) program repository, hence the need for very efficient similarity checks that will be iterated over all the programs in the repository; 
(3) The repository could contain similar but slightly different programs, due to variations in compilers or code versions.  Clone search must be robust to such variations; 
(4) Finally, the technique must work equally well on stripped binary codes (where symbols have been removed at compile time), handle the case where external function names are unavailable (for example IoT device firmware), and handle lightweight obfuscations (such as adding deadcode, or hiding literal identifiers).

All these constraints do not fit well with prior work on similarity, as state-of-the-art is increasingly focused on \textit{function-level}  similarities\footnote{According to Haq and Caballero~\cite{SBC},
since 2014, among 40 binary code similarity approaches, only 7  approaches have taken programs as input. 
}, with unclear scalability toward the program-level case.  
For example, we found in our experiments that
SMIT~\cite{LSMI} takes more than 43 hours to compute a similarity index between the main library of Geany and the cp command, while DeepBinDiff~\cite{DEEPBINDIFF}   is reported to take 10 minutes to compute  basic bloc matching on small binaries from the Coreutils package. 

\myparagraph{Goal} \textit{From the program clone search point of view, there is a strong need for a binary-level program-level similarity technique  that is {\it precise}, {\it robust} to slight variation, and {\it fast} enough  to operate over large code bases. This is exactly what we want to address in this paper.}  

\myparagraph{Our proposal} 
We explore  the application of \textit{spectral graph analysis}~\cite{chung1997spectral} to the problem of program clone search. It seems a very good starting point as, on graphs, it is both affordable and competitive against graph edit distances (GED) ~\cite{GED0} in terms of precision, while GED is arguably a very good (but expensive to compute) notion of graph similarity.  
Yet, programs are not standard graphs: on the one hand programs seen as graphs can be very large (especially at the binary level),  while on the other hand they are highly structured due to their function hierarchy.  

We take advantage of this specificity and  propose {\bf Program Spectral Similarity (\PSS)}, the first spectral analysis tailored to \textit{program} similarity.   
The techniques extract {\it eigenvalues} related features from both function call graphs and control flow graphs, and take advantage of a \textit{preprocessing} step (done once for the whole program repository) to achieve similarity checks in time \textit{linear in the number of functions of the program} (done for each program in the repository), making it a perfect fit for program clone search --  most prior works have at least a quadratic runtime.

We {experimentally} show that \PSS outperforms state-of-the-art approaches and  is resilient to code variations as well as lightweight obfuscations (e.g., instruction substitution, bogus control flow, control flow flattening). Moreover, \PSS does not rely on {\it literal identifiers} (e.g.,  function names, constant string values), making it robust against a range of basic obfuscations. 
In our experiments, a program clone search with \PSS (optimized version) takes on average less than 3s (0.3s and {0.4s} for Linux and IoT benchmarks) where,  
as a comparison, the function embedding Gemini~\cite{GEMINI} requires roughly 2 minutes per clone search.

We set up a strong comprehensive evaluation framework (14 competitors and 3 baselines) to systematically compare  \PSS with  state-of-the-art methods, covering string based methods~\cite{LibDB,LibDX}, graph edit distance~\cite{LSMI,GSA}, N-grams~\cite{MUTANTXS}, vector embedding~\cite{Asm2Vec,SAFE,GEMINI,AlphaDiff}, standard spectral methods~\cite{GSA}  and matching algorithms~\cite{SMM,MAV}. Our experiments cover our own dataset of diverse open-source projects along with classical Coreutils, Diffutils, Findutils,  and Binutils packages along two dimensions (optimization levels and code versions) for a total of 950 programs. 
Moreover, we consider part of the BinKit dataset ~\cite{kim:tse:2022} (98K samples),     
covering  four optimization levels, 9 compilers, 8 architectures and 4 obfuscations. 
Finally, we gather   $19,959$ IoT malware and  $84,992$  Windows goodware.

\myparagraph{Contribution} As a summary, we claim the following: 

\begin{itemize}
    \item A novel technique named \PSS (together with its optimization \PSSO)  for code similarity (Section~\ref{SectionPSS}), tailored to {\it program} clone search over {\it large} repositories. \PSS is the first  spectral technique tailored to program-level similarity. Especially, \PSS takes advantage of a preprocessing step to perform latter similarity checks in time linear w.r.t. the number of functions in the program, making it a perfect fit for program clone search over large repositories; %
    
    \item A comprehensive evaluation framework for program clone search (Section~\ref{XP}), encompassing (1)  97,760 programs from BinKit~\cite{kim:tse:2022}, 19,959 IoT malware,  84,992 Windows programs  and  a smaller Linux dataset of 950 programs, and (2)  three baselines and $14$ state-of-the-art methods -- $10$ of them being reimplemented. {\it The complete framework is available online}, which is rare in this field~\cite{MLFunctions};

    \item Experimental evidence (Sections~\ref{XP}) that 
    \PSS reaches a sweet spot in terms of speed, precision and robustness, making it a perfect fit for program clone search, where prior works in the field are more specialized to function-level similarity evaluation. Especially, \PSS appears to scale well and to retain good precision in demanding clone search scenarios (cross-compilers, cross-architecture or obfuscation);

    \item Finally, as another notable result, %
     we show that prior work targeting  function clones cannot cope with  program clones due to scalability issues.

\end{itemize}

Besides providing a novel and efficient method for program clone search, our results also shed new light on prior work on code similarity. 
First, we make the case for the program clone search application scenario and show that it behaves differently enough than the well-studied  pairwise function similarity setting, requiring dedicated methods. 
Second, we are the first to pinpoint the separation in prior work between techniques using literal identifiers and those that do not. As a side result, during our experiments, we identify two simple methods  based on literal identifiers (string values and external function names), which despite their simplicity, appear to perform well when these identifiers are available. These methods came from the simplification of ideas coming from the state-of-the-art in library identification by using literal identifiers~\cite{SMM, BAT, OSSPolice,LibDX}.
Third, we show the potential of dedicated spectral methods for program clone search. 
Overall, we believe that these results pave the way for novel  research directions in the field.

\smallskip 

{\bf Research artifacts are available on Zenodo~\cite{PSSOartifact}.}

\section{Problem statement}

\subsection{Program Clone Search Procedure}

Given an unknown \textit{target program} $P$ and a \textit{program repository} $R$, the goal is to identify a \textit{clone} of $P$ in $R$.

A clone of a program $P$ is defined as follows: 
\begin{itemize}
    \item A program $Q$ compiled from the same source code $S$ as $P$, but with a different compiler toolchain is a clone of $P$. For example, $P$ has been compiled with GCC v9.1 using the optimization level -O0 from the source code $S$, and $Q$ has also been built from $S$ using the same compiler but another optimization level, say -O3; 
    \item A program $Q$ compiled from another version of $P$ source code is a clone of $P$. For example, both instances of the git application compiled from two source code versions, say v2.35.2 and v2.37.1, are clones. 
\end{itemize}

\pagebreak

In the last case, we have to be a bit careful. Indeed, we can only consider incremental versions of an application or library, not major revisions that completely change the source code. In our experiments, the newest and oldest versions of most packages are usually separated by 4 years. However, it goes up to 15 years for the most standard packages: Coreutils, Diffutils, and Findutils.

\begin{figure}[htbp]
    \centering
    \includegraphics[width=1\columnwidth]{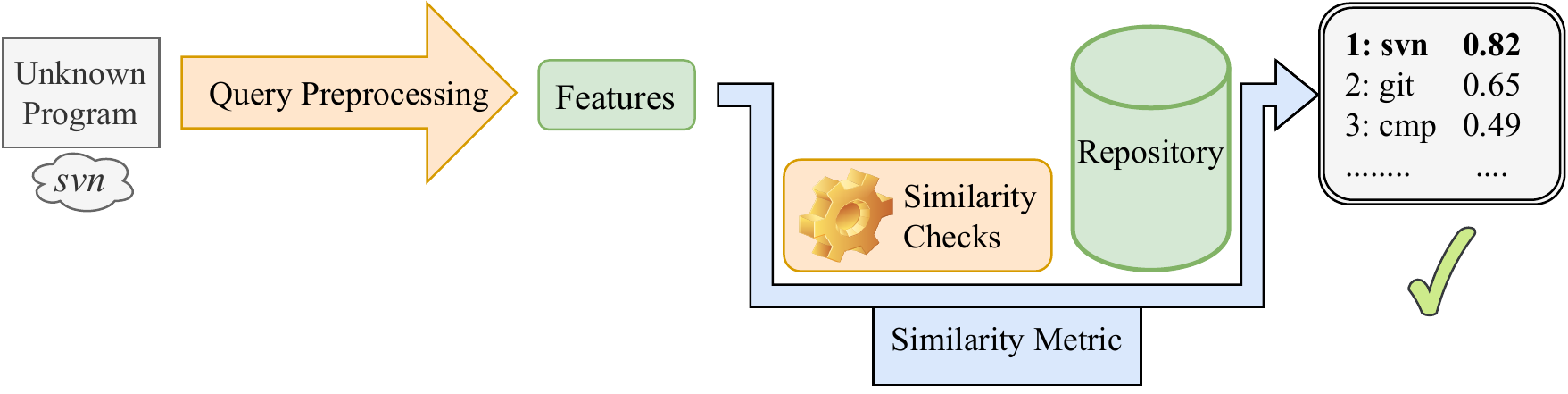}
    \caption{Architecture of a program clone search procedure}
    \label{fig:diagram}
\end{figure}

Figure~\ref{fig:diagram} illustrates a clone search procedure architecture. Note that all along, we suppose that there is no exact copy of $P$ in the repository $R$. 
The repository is a database containing enough information for a clone search procedure. As a result, in practice, a repository is quite an extensive program database w.r.t. the application domain (firmware, plagiarism, malware, etc.). 

An evaluation of clone search procedures should take into consideration the three criteria below in order to be realistic: 
 
\begin{itemize}
    \item The efficiency w.r.t. both the size of the unknown target program and the size of the repository, 
    
    \item The robustness not only to compiler toolchains but also to slight program variations coming from different source code versions, 

    \item The ability to deal with stripped programs. Moreover, external symbols are not necessarily available when dealing with firmware, lightweight obfuscations, or yet from payload extracted from packers\cite{PayloadPackers}.

 \end{itemize}

As we said previously, the main difference between program clone search and function clone search is the size of the binary codes, which is much larger in the case of programs.

At a high level, all program clone search procedures work in a similar way. The repository is already built, and the query process is divided into three steps: 
  
    \begin{enumerate}
        
        \item {\bf Query preprocessing.} Upon query, we receive the target program $P$. We can perform some preprocessing at this step, extracting relevant features for the rest of the procedure; 
        
        \item {\bf Similarity checks.} For each program $Q \in R$, we perform a similarity check with a similarity metric $M$ on $(P,Q)$ -- possibly taking advantage of the preprocessing --  and record the computed similarity index $M(P,Q)$; 
        
        \item {\bf Decision.} The program $Q_{best}$ with the highest similarity index is considered the most similar. The program clone search succeeds  if $Q_{best}$ is a clone of $P$,  otherwise  it fails.
    \end{enumerate}

\subsection{Motivating Example\label{motivatingInPractice}}
Let us consider a repository containing $1420$ libraries obtained from the compilation of $20$ libraries\footnote{From packages libiconv, coreutils, libtool, gss, gdbm, libtasn1, gsl, libmicrohttpd, osip, readline, gsasl, lightning, recutils, gmp, libunistring, and glpk.}  with  four optimization levels, five  versions of GCC, four versions of clang, and to the 32 and 64 bits x86 platforms. Next, let us imagine we have the 20 libraries as targets (compiled for x86 32 bits with gcc 6.4 and the -O2 optimization level).

Lifting function-level clone searches in order to detect program-level clones is attractive. However, to obtain a similarity index between two programs from function embedding methods,  %
we need to find a distance between two sets of function embeddings. Let $embeds(P)$ be the set of function embeddings of a program $P$. A first  solution is to perform a matching between the two sets. Such matching could be an instance of the assignment problem where assigning a function embedding $x$ of $P$ to a function embedding $y$ of $P'$ has a cost $\left\Vert x -  y\right\Vert_2$. However, this problem has complexity $O(n^3)$ where $n$ is the number of functions. We relax the matching so that a function embedding of a program $P$ can be assigned to multiple function embeddings of a program $P'$.

We define $F$ as the similarity metric for an embedding $embeds$:
\begin{align}
F(P,P') := -\sum_{x \in embeds(P)} \min_{y \in embeds(P')}  \left\Vert x -  y\right\Vert_2 \label{functionadaptation}
\end{align}

We consider the following  function-level methods and lift them to programs as just explained: {\it Asm2Vec}~\cite{Asm2Vec},  {\it Gemini}~\cite{GEMINI}, {\it SAFE}~\cite{SAFE},  {\it $\alpha$Diff}~\cite{AlphaDiff}.
We also consider {\it LibDB}~\cite{LibDB}, which is directly designed for libraries (i.e., large pieces of code).

\begin{table} 
\caption{Clone searches results}
\label{tab:cloneexample}
\begin{tabular}{|l|c|c|c|c|}
\hline
Framework  & Average & Total runtime \\
  & precision@1  &  (preprocess. time included) \\
\hline
{\it Asm2Vec}~\cite{Asm2Vec} $\dagger$  & 0.7 & 35h   \\ %
{\it Gemini}~\cite{GEMINI} $\dagger$  & 1 & 17h   \\  %
{\it SAFE}~\cite{SAFE} $\dagger$  & 0.95 & 160h  \\ %
{\it $\alpha$Diff}~\cite{AlphaDiff} $\dagger$  &  1 & 140h \\ %

{\it LibDB}~\cite{LibDB} $\dagger$  & 1 & 2h \\ %
\hline
\hline 
\textbf{\PSS} & 1 &  \textbf{26s}  \\
              &   &   (includ. 26s of preprocess)  \\
\hline
\end{tabular}

$\dagger$ learning time not included

\end{table}

\myparagraph{Results} We report in Table~\ref{tab:cloneexample} the average precision@1, equivalent to the proportion of successful clone searches, as well as clone searches total  runtime.
\PSS is precise and successful in all clone searches. Most function-level methods can also find a clone in all clone searches. However, \PSS takes only 26s in total, while pure function embedding methods take from 17h with {\it Gemini} to 160h with {\it SAFE}. Even with pre-filtering, {\it LibDB} is close to 2h. Moreover, \PSS runtime is due to its preprocessing; the total similarity checks runtime is negligible. As a result,  \PSS scales up to large repositories with good precision.

\pagebreak

\section{Background}

\myparagraph{Graph similarity, GED and spectral distance} As programs can be naturally seen as graphs, any good notion of graph similarity is in principle a good candidate for a good program similarity metric.  

Graph edit distance (GED) is such a good notion~\cite{SurveyGED}. GED is the smallest cost of an edit path between two graphs, i.e. the smallest transformation going from one of the graphs to the other. Graph edit operations typically include removing or adding a vertex or an edge.
Yet, the main drawback of GED is that its computation is NP-hard.
Worst, usual approximations  have a complexity of $O(n^3)$~\cite{SERRATOSA2014244}  where $n$ is the number of nodes in the graph, which is far too expensive for graphs coming from programs. As an example, the graph edit distance method {\it SMIT}~\cite{LSMI} is the slowest method we have tested  (cf.~Table~\ref{tab:basictotalrunning}), with 3634 hours of computation on a task where our method takes 1h18m. 

The spectral distance between graphs provides an interesting trade-off, as it gives a  decent approximation of the graph edit distance between graphs~\cite{SGSCG} for an affordable linear cost once eigenvalues are computed. We introduce spectral analysis and define spectral distance hereafter.

\myparagraph{Spectral (Graph) Analysis} 
Spectral graph analysis is a method used to investigate properties of graphs by studying the eigenvalues (or, \textit{spectrum}) of standard matrices associated with the graph, such as the adjacency matrix or the Laplacian matrix. Patterns and structures within the graph can be identified, providing key insights about how the graph nodes are interconnected. Distances between graph spectra are called spectral distances.
The starting intuition for using graph spectrum is that two isomorphic graphs have the same spectrum; however, the converse is not true. Nevertheless, the spectrum may be used as a proxy for graph similarities. 

More formally, an undirected graph $G=(V,E)$ of $n$ vertices is represented by an $n \times n$ adjacency matrix $A$, where $a_{i,j}$ is one if $(V_i,V_j) \in E$ and zero otherwise. Let $d_i$ be the degree of the vertex $V_i$. 
It is useful to compute the Laplacian matrix~\cite{chung1997spectral}  $L$ of G.
An eigenvalue $\lambda$ and its corresponding eigenvector $\vec{u}$ is a solution to the equation: 
 $ \left(L - \lambda I \right) \vec{u} = \vec{0} $.
The spectrum is the set $\{\lambda_1(G),\ldots,\lambda_{|G|}(G)\}$
where $\lambda_1(G) \geq \ldots \geq \lambda_{|G|}(G)$
and where $|G|$ is the number of vertices in $G$. The enhanced Lanczos algorithm~\cite{Lanczos} computes the spectrum in time $O(dn^2)$, where $d$ is the average degree of $G$. 
We define the spectral distance between $G_1$ and $G_2$ (analogous to\cite{RefSpectralDistance}):

$
sD(G_1,G_2) := \sqrt{\sum_{i=1}^{min(|G_1|,|G_2|)}   \left(\lambda_i(G_1) - \lambda_i(G_2)\right)^{2}}
$.

\section{Program Spectral Similarity (PSS)\label{SectionPSS}}
Spectral analysis is suitable for comparing graphs because it provides quantitative metrics, such as spectral distances, which can be used to compare key graph properties regarding connectivity, structure, and distribution. This approach also allows for the normalization of graph size, enabling fair comparisons among varying graph scales.
However, computing the spectrum of a graph is cubic in its number of nodes. Therefore, applying spectral analysis to a whole program CFG is too expensive. Moreover, the CFG itself is not stable with respect to compiler toolchains, optimizations and obfuscations.

As a result, our key insight is that a program has more structure than a mere graph: there is a call graph over functions while local functions hold their own control flow graph. We take advantage of this hierarchical structure to devise a quick and stable similarity metric called \textit{Program Spectral Similarity} (\PSS).

The \PSS method is based on the combination of two criteria. 
\begin{itemize}
\item The first measure is the spectral distance between call graphs, including both internal and external calls\footnote{Call graphs are useful for a number of tasks. For example, GraphEvo~\cite{GraphEvo} has been able to understand software evolution through call graphs.}. Moreover, most compiler optimizations have a small-scale effect on the call graph, as they only impact the content of functions;

\item The second measure is a coarse spectral analysis of function control flow graphs, simply  considering their number of edges, as it is related to the sum of the eigenvalues as shown below. Since we use only one number to represent a function CFG, we can fit these numbers into a vector comparable to the eigenvalues vectors. Adding function embeddings to the second measure is left for further work.

\end{itemize}

By the way, we tried to consider only the control flow graphs, and we found that the results were worse than when both above criteria were considered. 

The \PSS method proceeds into two independent steps: the preprocessing step, which is done once and for all, and the similarity check step, which is made for each {candidate}.

\subsection{Preprocessing\label{preprocessingspp}}

\begin{figure}[htbp]
    \centering
    \includegraphics[width=0.475\textwidth]{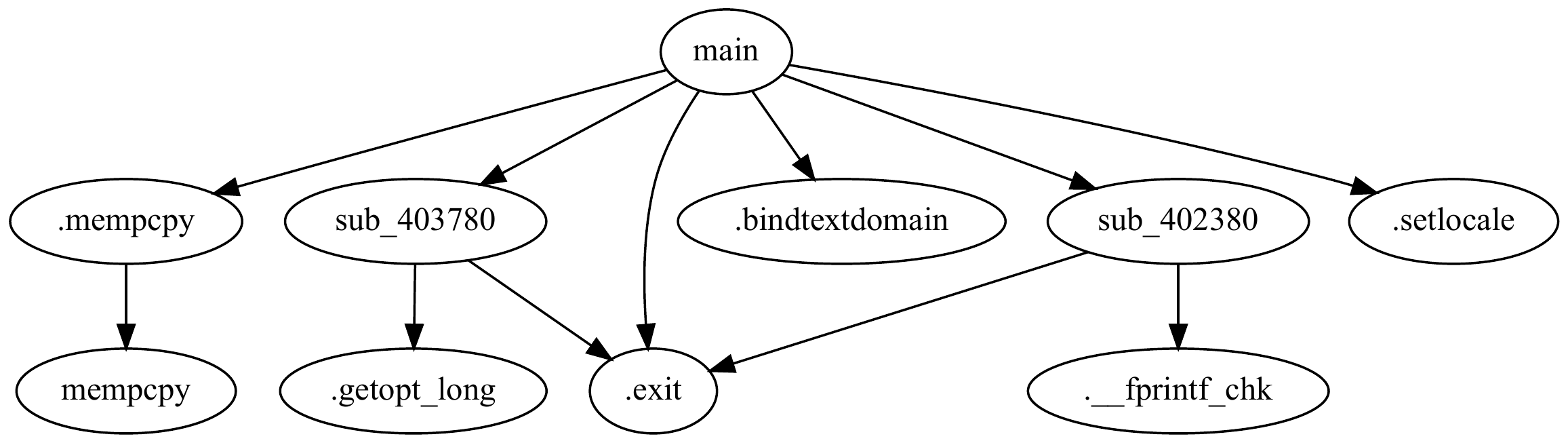}
   
    \caption{A call graph}
    \label{fig:FCG}%
\end{figure}

Given a program $P$, the preprocessing first begins by building the function call graph $CG$ of $P$, including local and external (API) calls. An example of a function call graph is given in Figure~\ref{fig:FCG}. It contains external calls such as a call to \texttt{mempcpy} as well as local functions such as \texttt{sub\_403780}. 
From this, we extract two key vector-features $(\vec{v},\vec{w})$ of $P$ as follows:

\begin{itemize}
\item From an undirected version of the call graph $CG$, we compute the spectrum $\Lambda = \{\lambda_1(CG),\ldots,\lambda_n(CG)\}$, and we compute $\vec{v} := \frac{\Lambda}{\|\Lambda\|_{2}}$, the normalized spectrum of the call graph;

\item We compute the number of edges $E=(e_1,e_2,\ldots,e_k)$ from each control flow graph $F_i$ of local functions in descending order, %
and we normalize $E$ as previously: $\vec{w} := \frac{E}{\|E\|_{2}}$. External functions are ignored at this step since we do not have access to their control flow.
Note also that the number of edges is a simple sort of  spectral measure since it is related to the spectrum by the relation $2 \times e_i = \sum \lambda_j(F_i)$. 

\end{itemize}

Recall that $||\cdot||_2$ is the Euclidean norm. We normalize features $\vec{v}$ and $\vec{w}$ to deal with differences between program sizes.

\subsection{Similarity Check\label{checkspp}}
Given two programs, $P_0$ and $P_1$, the preprocessing step has computed features $(\vec{v_0},\vec{w_0})$ from $P_0$, and $(\vec{v_1},\vec{w_1})$ from $P_1$. 
The similarity check outputs a similarity index by averaging two measures. The first measure (\ref{EQ:SIMCG}) is related to call graphs, while the second (\ref{EQ:SIMCFG}) is related to function control flow graphs. Then, the similarity metric $\PSS$  is defined as the average of both above measures (Equation~\ref{EQ:PSS}).

\begin{align}
\label{EQ:SIMCG}
simCG(P_0,P_1) := \sqrt{2} - \sqrt{\sum_{i=0}^{min(|\vec{v_0}|,|\vec{v_1}|)}   \left(v_{0,i} - v_{1,i}\right)^{2}}
\end{align}
\begin{align}
\label{EQ:SIMCFG}
simCFG(P_0,P_1) := \sqrt{2} - \sqrt{\sum_{i=0}^{min(|\vec{w_0}|,|\vec{w_1}|)}   \left(w_{0,i} - w_{1,i}\right)^{2}}
\end{align}
\begin{align}
\label{EQ:PSS}
\PSS(P_0,P_1) :=  \frac{simCG(P_0,P_1) + simCFG(P_0,P_1)}{2\sqrt{2}}
\end{align}

\subsection{The PSSO Optimization}
We found out that \PSS preprocessing may be quite long over large programs (cf.~our own "Windows dataset" in Section~\ref{RQSpeed}, where computing all eigenvalues of a call graph takes $16.95$ seconds per program clone search.).
In order to tackle this issue, instead of computing the complete spectrum $\Lambda$, we propose to compute only the first $K$ greater eigenvalues so that  $ \Lambda = \{\lambda_1(CG),\ldots,\lambda_{K}(CG)\}$. For this, we can take advantage of a variant of the Lanczos algorithm proposed by the ARPACK library~\cite{ARPACK}.

\begin{figure}[htbp]
    \centering
    \includegraphics[width=1\columnwidth]{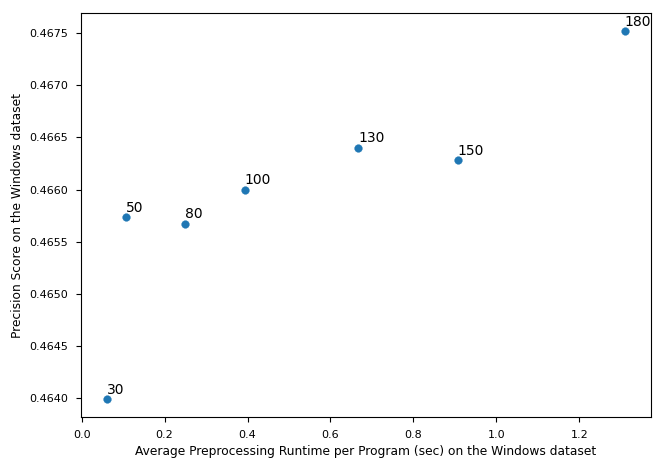}
    \caption{Impact on the Windows dataset of the number of largest eigenvalues computed by PSS optimized version}
    \label{fig:plotwindows}
\end{figure}

We plot in Figure~\ref{fig:plotwindows} the preprocessing runtimes and precision scores (see Section~\ref{sec:methodo}) for different values of $K$ from 30 to 180 on our "Windows data set". We remark that runtimes grow quickly with $K$, going from $0.06$s to $1.31$s. On the other hand, 
there is little change in the precision score between 50 and 150; the score varies from $0.4657$ to $0.4664$. We select $100$ as the value for $K$ since the preprocessing runtime per clone search is only $0.39$s, and the precision score is already $0.4661$.

We thus propose \PSSO, an optimized version of \PSS that computes only the first $K=100$ greater eigenvalues.

\subsection{Method Runtimes}
\begin{table}[htbp]
\centering
\caption{Complexity of program clone search procedures}

\begin{tabular}{|l|l|l|l|}
\hline
Method  & Class        & Similarity &  Preprocess.$\ddagger$  \\
           &              & check$\dagger$     &  \\    

\hline
{\it SMIT}~\cite{LSMI}  & GED    & $O(n^4)$  & $O(dn)$ \\
\hline
{\it CGC}~\cite{SMM} & Matching   & $O(n^4)$ & $O(dn)$ \\
\hline
\textbf{\textit{MutantX-S}}~\cite{MUTANTXS} & {\bf N-gram}   &  {\bf $O(1)$} &  $O(i)$\\
\hline
{\it Asm2Vec}~\cite{Asm2Vec} & Functions ML & $O(n^2)$ & $O(n)$ \\
{\it Gemini}~\cite{GEMINI} & Functions ML   & $O(n^2)$ & $O(n)$  \\
{\it SAFE}~\cite{SAFE} & Functions ML   & $O(n^2)$ & $O(n)$ \\\hline
{\it $\alpha$Diff}~\cite{AlphaDiff} & Functions ML   & $O(n^2)$  &  $O(n)$ \\
\hline
\textit{LibDX}~\cite{LibDX}  & Strings  & $O(s)$ & $O(s)$   \\
\textit{LibDB}~\cite{LibDB}  & Functions ML  & $O(n^2 + s)$   &   $O(n + s)$                             \\
  & and Strings  &    &               \\
\hline
{\it DeepBinDiff}~\cite{DEEPBINDIFF} & ML   & $O(n^{3}m^{3})$  & no preproc. \\

\hline
\hline
\textbf{\textit{PSS}}  & {\bf Spectral}     &  {\bf $O(n)$} & $O(dn^2)$ \\
\textbf{\textit{PSS$_O$}}  & {\bf Spectral}     &  {\bf $O(n)$} & $O(dn)$ \\

\hline
\end{tabular}
\medskip

$n$: \# functions, $i$: \# instructions , $s$: \# constant string values

$d$: \# calls per function, $m$:  \# basic blocks in a function,

$\dagger$  between two programs

$\ddagger$ performed once for the whole clone search

\label{tab:methodsIntro}

\end{table}

Recall that a repository is a database of preprocessed programs. A given unknown target program is first preprocessed,  then, from the extracted features, a similarity check  is made on the repository. It is clear that the query runtime linearly depends on the size of the repository.
In other words, for a repository size of $M$, and  $n$ the number of functions inside a program, if the runtime of a similarity check is $T(n)$ and the preprocessing runtime is $P_T(n)$,  then the complexity of a query is bounded by $M \times T(n) + P_T(n)$. As a result, all methods with similarity checks with superlinear time complexity are not feasible over large repositories of large codes, which is confirmed by our experiments. 

\myparagraph{\PSS and \PSSO runtimes} Graphs and Laplacian matrices are sparse in our application domain, offering quick eigenvalues computation.  Nevertheless, the complexity of the query prepossessing, described in Section~\ref{preprocessingspp}, is still $O(dn^{2})$, where $n$ is the number of functions and $d$ is the average number of calls per function. However, once such prepossessing is done, the runtime of a similarity check, described in Section~\ref{checkspp}, is $O(n)$. Moreover, the runtime of the query preprocessing of  \PSSO is reduced to $O(dn)$. 

\myparagraph{Comparison with prior work} That is in contrast with function embedding methods which have a similarity check runtime of $O(n^2)$ on this problem using a direct adaptation (see Section~\ref{motivatingInPractice} for further details). Moreover, {\it DeepBinDiff}~\cite{DEEPBINDIFF} contains a step with a linear assignment between basic blocs with a runtime of $O(n^{3}m^{3})$. Worse, both the  graph edit distance approximation {\it SMIT}~\cite{LSMI} and the matching method of Xu et al.~\cite{SMM} have a complexity of $O(n^4)$. However, the runtime of {\it MutantX-S}~\cite{MUTANTXS}, designed to scale up to large repositories, is only $O(1)$ -- yet experiments (Tables~\ref{tab:scoreshighlight} and~\ref{tab:correlations} in Section~\ref{RQRobustness})  %
show that its 
{robustness} %
 is not fully satisfactory.

\section{Systematic evaluation \label{XP}}
We  evaluate the potential of \PSS  in terms of speed, precision and robustness -- the ability to overcome changes in  compilation.

\noindent \textbf{Then, we consider here the following Research Questions:}
\begin{enumerate}
\item[RQ1] What are the fastest methods for clone search?
\item[RQ2] What are the most precise methods for clone search?
\item[RQ3] What are the most robust methods for clone search?
\item[RQ4] What is the impact of each component of PSS?
\end{enumerate}

\subsection{Datasets}

\myparagraph{Basic dataset}
We first collect a limited dataset of $950$ programs to study the full range of methods along different optimization levels and code versions. The average program has a size of \SI{442}{KB}. This dataset covers the Coreutils, Diffutils, and Findutils packages compiled with GCC v5.4 on the x86 architecture, and taken from the DeepBinDiff~\cite{DEEPBINDIFF} dataset. Moreover, we add the Binutils package as well as  {15}
open-source projects~\label{bigprojects}, including  {Bash}, {{Code::Blocks}}, {Dia}, {Graphviz}, {Geany}, {Git}, {Lua}, {Make}, {OpenSSH}, {OpenSSL}, {Perl}, {Ruby}, {SDL}, {SVN}, and {VLC}, 
compiled by GCC v9.4 on an x86 architecture. Each unique source code comes in four different {\it version levels}, and four different {\it optimization levels}. These programs are all clones of each other.

\myparagraph{BinKit dataset} To study scalable methods along different optimization levels, compilers, architectures, and obfuscations, we reuse two Linux programs datasets from BinKit~\cite{kim:tse:2022}: 

\begin{itemize}
  \item {\bf Normal:}  From 51 GNU software packages, 235 unique source codes were extracted. They are compiled with 288 different toolchains for a total of 67,680 programs of an average size of \SI{201}{KB}. It covers eight architectures (arm, x86, mips, and mipseb, each available in 32 and 64 bits), nine compilers (five versions of GCC and four versions of Clang), and the four optimization levels from -O0 to -O3; 
  \item {\bf Obfuscation:} Four obfuscation options (instruction substitution (SUB), bogus control flow (BCF), control flow flattening (FLA), and all combined) are considered using Obfuscator-LLVM~\cite{ObfuscatorLLVM} as a compiler. The same architectures and optimization levels as before are covered, for a total of 30,080 programs of an average size of \SI{514}{KB}.
\end{itemize}

\myparagraph{IoT Malware dataset} We consider 19,959 IoT malware samples, with an average size of \SI{84}{KB}, from MalwareBazaar\footnote{\url{https://bazaar.abuse.ch}}, submitted between March 2020 and May 2022, spanning  8 architectures (mostly arm, mips, motorola and sparc). Using available meta-data from antivirus reports and YARA rules, we split the data into only three families of clones: 12,357 Mirai, 5,842 Gafgyt, and 1,760 Tsunami. 

\myparagraph{Windows dataset}\label{section:windowsdataset} We assemble a dataset of 84,992 benign programs running under Windows operating systems (x86, Visual Studio). This amounts to more than \SI{50}{GB} of raw programs, with an average size of \SI{771}{KB}.
Excluding security updates, the dataset contains more than 28,000 dynamic-link libraries. Samples are divided by target platforms (e.g., Windows 7). 
We consider that two programs sharing the same file name and the same target platform are clones, yielding 49,443 programs with a clone. %

\subsection{Methodology} \label{sec:methodo}
A {\it test field} $(T,R)$ comprises targets set $T$ and a repository $R$.
We break down  Basic datasets along version levels and optimization levels. For instance, the test field (-O0,-O1) of the subdataset "Coreutils Option"  consists of a repository of Coreutils programs compiled with -O1 paired with the same programs but compiled with -O0 as targets. Similarly, we break down BinKit datasets along optimization levels, compilers, architectures, and obfuscations.

\myparagraph{Measures of success: precision@1} 
Program clone search is an {\it information retrieval}  task. The standard evaluation metrics of information retrieval are precision and recall. 
This study uses the evaluation metric described in the Asm2Vec paper~\cite{Asm2Vec}, that is Precision at Position 1 (precision@1). Precision@1 is equal to one if and only if a clone of the target is the most similar program in the repository, as ranked by a similarity metric. We define the precision {\it score} of a similarity metric as the average precision@1 for every target in every test field against a repository.

\subsection{Competitors}

We evaluate 14 competitors, 3 baselines  and two new heuristics based on literal identifiers (constant string values and external function names) (cf.~Table~\ref{tab:methods}). 
8 of these frameworks have been adapted {\bf (A)} to the case of program clone search, as it was not their primary objective (e.g., function embedding). 
Moreover, 10 had to be reimplemented {\bf (R)} because the original implementation was unavailable or due to inherent challenges in effectively utilizing the original implementation within the specific domain of clone search.
As highlighted by Marceli et al.~\cite{MLFunctions}, code similarity artifacts are rarely available, and even when they are, they are often incomplete.

\myparagraph{Baseline} We first investigate basic heuristics such as \textit{$B_{size}$}, the size of the program, and \textit{$D_{size}$}, the size of the disassembled program. For instance, the similarity metric \textit{$B_{size}$} is defined as $B_{size}(a,b) := -\lvert a - b \rvert$, where $a$ and $b$ are program sizes in bits. We also consider a crude shape of the call graph. Let $n_1$ and $e_1$ (respectively $n_2$ and $e_2$) be the number of vertices and edges of the first (respectively second) call graph. Then the similarity measure \textit{Shape} is defined as: 
$$Shape(n_1,e_1,n_2,e_2) := \frac{\min(n_1,n_2)}{\max(n_1,n_2)}\times\frac{\min(m_1,m_2)}{\max(m_1,m_2)}$$

\myparagraph{Standard spectral methods} From the spectral method developed by Fyrbiak et al.~\cite{GSA},  we derive two methods. The first, \textit{ASCG} {\bf (A)  (R)}, is based on the call graph. Let $X$ and $Y$ be the two spectrums in descending order of Laplacians of the two call graphs. There is a normalization $X' := X/X_0$ and $Y' := Y/Y_0$. Then:  $$ASCG(X', Y') :=  -\sum_{i=0}^{min(|X'|,|Y'|)} \left | X'_{i} - Y'_{i} \right|  $$

Likewise, we derive a method based on the control flow graph,  \textit{ASCFG} {\bf (A) (R)}. Instead of computing the spectrum from the call graph, we select the top 1000  eigenvalues from a reduced control flow graph as vectors $X$ and $Y$.

\myparagraph{Graph edit distance} We implement various basic GED based methods. First, we implement \textit{GED-0} {\bf (A) (R)}, a basic computation of the GED applied between call graphs. The algorithm goes back to the work of Sanfeliu and Fu~\cite{GED0}.   Second, we implement \textit{GED-L} {\bf (A) (R)}, a computation of the GED between call graphs with labels. The algorithm is presented by Fyrbiak et al.~\cite{GSA}. In our application, labels are sets of external function names. Third, we implement the specific GED computation of Hu et al.~\cite{LSMI} called \textit{SMIT} {\bf (R)}. We do not integrate the indexing tree of SMIT as we are more interested in their GED measure.

\myparagraph{Matchings}  We compare with the matching algorithm {\it CGC} {\bf (R)} from Xu et al.~\cite{SMM}. This algorithm needs three parameters along with a complete classification of mnemonics. We perform preliminary works to find good values for these parameters. 

\begin{table}[htbp]
\centering
\caption{Methods included in the evaluation}

\begin{tabular}{|l|l|l|l|l|l|}
\hline
Framework  & Class  & A & R & Similarity & LIR \\
  &         &   & & check                    &  \\
\hline

\textbf{\textit{$B_{size}$}} & \textbf{\textit{Baseline}}    & &  & \textbf{\textit{$O(1)$}}                    &                         \\
\textbf{\textit{$D_{size}$}}  & \textbf{\textit{Baseline}}   & & & \textbf{\textit{$O(1)$}}                    &                         \\
\textbf{\textit{Shape}}  & \textbf{\textit{Baseline}}  & &  & \textbf{\textit{$O(1)$}}                    &                          \\ \hline

\textbf{\textit{ASCG}} ~\cite{GSA}   & \textbf{\textit{Spectral}}  &  \Cross  &  \Cross  & \textbf{\textit{$O(n)$}}                    &                           \\
\textbf{\textit{ASCFG}} ~\cite{GSA}   & \textbf{\textit{Spectral}} &  \Cross &  \Cross   & \textbf{\textit{$O(1)$}}                    &                           \\
{\it GED-0}~\cite{GED0} & GED   & \Cross  &  \Cross  & $O(n^3)$ &                  \\
\textbf{\textit{MutantX-S}}
\cite{MUTANTXS} & 
\textbf{N-gram} & &  \Cross  & \textbf{$O(1)$} &   \\
{\it Asm2Vec}~\cite{Asm2Vec} & Function ML &  \Cross &   & $O(n^2)$ &                \\
{\it Gemini}~\cite{GEMINI} & Function ML  &  \Cross &  & $O(n^2)$ &                    \\
{\it SAFE}~\cite{SAFE} & Function ML  &  \Cross &  & $O(n^2)$ &  \\
{\it DeepBinDiff}~\cite{DEEPBINDIFF}& ML   & & & $O(n^{3}m^{3})$ &                 \\
\hline
\textbf{\textit{PSS}} & \textbf{\textit{Spectral}}     &  & &   \textbf{\textit{$O(n)$}}   &                  \\
\textbf{\textit{PSS$_O$}}   & \textbf{\textit{Spectral}}  &  & &   \textbf{\textit{$O(n)$}}     & \\
\hline
\hline

{\it GED-L}~\cite{GSA} & GED & \Cross &  \Cross    & $O(n^3)$ &    \Cross  \\
{\it SMIT}~\cite{LSMI} & GED  & &  \Cross   & $O(n^4)$ &    \Cross   \\
{\it CGC}~\cite{SMM} & Matching  & &  \Cross & $O(n^4)$ &    \Cross  \\
{\it $\alpha$Diff}~\cite{AlphaDiff} & Function ML &  \Cross &  \Cross  & $O(n^2)$ &   \Cross     \\
\textit{LibDX}~\cite{LibDX}  & Strings  & & \Cross & $O(s)$                    &    \Cross           \\
\textit{LibDB}~\cite{LibDB}  & Strings and & & \Cross &   $O(n^2 + s)$                  &    \Cross           \\
  & Function ML  & &  &                   &               \\
\hline
\textit{StringSet}  & Strings  & &  & $O(s)$                    &    \Cross           \\

\textbf{\textit{FunctionSet}}  & \textbf{\textit{Strings}}  & &  & \textbf{\textit{$O(n)$}}                    &    \Cross           \\
\hline
\end{tabular}
\medskip

A: Adapted for program clone search, R: Reimplemented

LIR: Some literal identifiers are required

\label{tab:methods}

\end{table}

\myparagraph{N-gram} We reproduce {\it MutantX-S}  {\bf (R)} from the work of Hu et al.~\cite{MUTANTXS}. We extended it to multiple architectures. Each program is represented by the frequencies of 4-grams obtained from the opcode sequence. These frequencies are embedded into a 4096-dimension vector by hashing.

\myparagraph{Function embeddings} As previously, we use the similarity metric $F$ to compare sets of vector embeddings (refer to Equation~\ref{functionadaptation} in Section~\ref{motivatingInPractice}).
We first consider {\it Asm2Vec} {\bf (A)}~\cite{Asm2Vec}. We employ an unsupervised training strategy on the Basic dataset inspired by the original paper. Multiple training phases are performed, with each time one optimization level for training and one for testing.
Then, we take  {\it Gemini} {\bf (A)} embedding from Xu et al.~\cite{GEMINI} in an optimistic setting. We build a version of the basic dataset retaining function names and employ these as ground truths for training. 
Moreover, we use the embedding of Massarelli et al.~\cite{SAFE} with  {\it SAFE} {\bf (A)}. We downloaded a pre-trained model made available by one of the authors\footnote{\url{https://github.com/facebookresearch/SAFEtorch}}.
Lastly, we reproduce {\it $\alpha$Diff} {\bf (A) (R)} from the framework of Liu et al.~\cite{AlphaDiff}. It is tailored to binary function similarity between versions. We sample 25\% of the $\alpha$Diff dataset\footnote{\url{https://twelveand0.github.io/AlphaDiff-ASE2018-Appendix}}  as our training set. {\it $\alpha$Diff} incorporates external function names and in-out degrees in the call graphs.

\myparagraph{DeepBinDiff} The framework {\it DeepBinDiff} from Duan et al.~\cite{DEEPBINDIFF} attempts to match basic blocs between two binaries. The similarity metric computes the number of matched basic blocs by {\it DeepBinDiff} between two programs. Due to its runtime, we were unable to perform experiments, and it is only considered inside the preliminary evaluation.

\myparagraph{LibDX} We reproduce the framework {\it LibDX}  {\bf (R)} from Kim et al.~\cite{LibDX}. It extracts constant string values from well-defined read-only sections of programs. Constant string values are compared with matchings and the tf–idf statistic.

\myparagraph{LibDB} We  reproduce the framework {\it LibDB} {\bf (R)} from Kim et al.~\cite{LibDB}. They combine function embeddings and matchings, while using constant string values as pre-filters. 
We reimplemented {\it LibDB} with our trained {\it Gemini} model  and ScaNN~\cite{scann} as the nearest vector search engine.

\myparagraph{Function set method} Xu et al.~\cite{SMM} describe a simple method that first matches functions between two programs by using only external function names and mnemonics similarities. Then, the similarity measure is computed by a distance over the two function sets. We simplify this idea and invent the similarity metric \textit{FunctionSet}, which computes the Jaccard similarity index\footnote{\url{https://en.wikipedia.org/wiki/Jaccard_index}} between external function names. Let $F_a$ be the external function names set of a program $a$. The similarity metric is: 
$FunctionSet(a,b):= \frac{|F_a \cap F_b|}{|F_a \cup F_b|}$.

\myparagraph{String set method \label{ref:fsss}} We invent a straightforward metric that compares constant string values inside programs.  Let $S_a$ be the set of all constant string values of a program $a$. The similarity metric is: $StringSet(a,b):= \frac{|S_a \cap S_b|}{|S_a \cup S_b|}$.

\smallskip

We present in Table~\ref{tab:methods} the characteristics of the different methods considered here. We record  the  runtime complexity of a similarity check between two programs. We note with $n$, $m$, and $s$, the number of functions, basic blocs in a function CFG, and literal identifiers respectively. We indicate whether a method requires literal identifiers. Note that machine learning approaches require a learning phase, and {\it Gemini} and {\it GCG} require manual mnemonics classification.

\myparagraph{Implementation} Disassembly is implemented by running the IDA Pro disassembler v7.5 along with a script from the Kam1n0 assembly analysis platform\footnote{\url{https://github.com/McGill-DMaS/Kam1n0-Community}}. 
See the recent survey of Pang et al.~\cite{SOK} on disassembling for more details. Our experiments are run on a cloud server node containing two CPUs  with a frequency of 2.10 GHz and 20 cores per CPU. 
 {All reported runtimes are equivalent to runtimes using only one core.}

\begin{table}[!htb]
    \caption{(RQ0) Total runtimes on the Basic dataset}
    \label{tab:runtimesbasic}
    \label{tab:basictotalrunning}
    \begin{minipage}{.5\linewidth}
      \centering
\begin{tabular}{|l|c|}
\hline
\textbf{\textit{B$_{size}$}} & $\leq$ {1h30m}
   \\
\textbf{\textit{D$_{size}$}}  & $\leq$ {1h30m}
    \\
\textbf{\textit{Shape}}           &                $\leq$ {1h30m}
   \\
\textbf{\textit{ASCG}} &           $\leq$ {1h30m}
      \\
\textbf{\textit{MutantX-S}} & $\leq$ {1h30m}
    \\
\hline
\textbf{\textit{PSS}}  & $\leq$ {1h30m}
   \\
\textbf{\textit{PSS$_O$}}  & $\leq$ {1h30m}
   \\
\hline 
\textbf{\textit{LibDX}} & $\leq$ {1h30m}
   \\
\textbf{\textit{StringSet}} & $\leq$ {1h30m}
     \\
\textbf{\textit{FunctionSet}}  & $\leq$ {1h30m}
    \\
\hline
\end{tabular}
    \end{minipage}%
    \begin{minipage}{.5\linewidth}
      \centering

\begin{tabular}{|l|c|}
\hline
{\it ASCFG} &  {128h} \\ %
{\it GED-0} &      81h \\ %
{\it GED-L}  & 46h \\ %
{\it SMIT}  &  3634h \\ %
{\it CGC} & 171h \\ %
{\it Asm2vec} $\ddagger$ & 141h \\ %
{\it Gemini} $\ddagger$ & 102h \\ %
{\it SAFE} $\ddagger$ &   655h\\ %
{\it $\alpha$Diff}  $\ddagger$  &    642h \\ %
{\it LibDB} $\ddagger$ & 16h \\ %
\hline
\end{tabular}
\end{minipage} 
\medskip

{\bf fast methods selected for further analysis}

$\ddagger$:  Learning time not included
\end{table}

\pagebreak

\subsection{Preliminary Evaluation: Method Selection}
\label{RQSpeedElimination}

{\it First, we want to identify methods unable to scale to large benchmarks, in order to not consider them in further analysis. We perform  a speed assessment on the basic dataset of 950 programs, and remove the methods unable to achieve it in less than {1h30m}.} 

\myparagraph{Results} 
Results are presented in Table~\ref{tab:runtimesbasic}. 
Please note that we could not experiment on  DeepBinDiff~\cite{DEEPBINDIFF} (with an observed average of more than 10 minutes per similarity check,  we estimate that it would have taken more than 20,000h to  apply it to the whole basic benchmark), and the training time of ML based methods is not counted in the reported timing.  
Results show a significant dichotomy between methods, 10 of them being able to succeed in less than {1h30m}
(often far less), while the other ten methods require far more time (from 16h to 3634h).

\myparagraph{Conclusion}  This preliminary experiment shows that  function-level clone search methods (typically based on ML)~\cite{Asm2Vec,GEMINI,SAFE,AlphaDiff,LibDB}  or graph-edit distance approaches~\cite{GED0, GSA, LSMI} cannot scale to program-level clone search. In the following, we will consider only the scalable-enough methods, namely our three baselines ($B_{size}$, $D_{size}$, {\it Shape}), as well as {\it ASCG}~\cite{GSA}, {\it MutantX-S}~\cite{MUTANTXS},  {\it LibDX}~\cite{LibDB} and our own {\it PSS}, {\it PSS$_O$}, {\it StringSet} and {\it FunctionSet}.

\begin{table}
\centering
\caption{(RQ1) Total runtimes. Include preprocessing time. Significant preprocessing times reported in "( \ )".}
\label{runtimesTotal}
\begin{tabular}{|l|c|c|c|c|}
\hline
Dataset &  Basic & BinKit & IoT & Windows \\
{\# Programs} &   1K & 98K & 20K & 85K \\
\hline
\textit{B$_{size}$} & 6s   & 43h & 	47m &  8h41m \\
\textit{D$_{size}$}  & 5s  & 43h & 	47m &  8h45m \\
\textit{Shape} & 1m22s  &  21h25m  & 21m26s &  4h16m \\
\hline
\textit{ASCG}  & {1h18m}%
& 143h & 1h23m  &  {243h} \\ %
\qquad preproc. &  {(1h18m)}%
&  (81h) &   (19m12s) &  {(228h)} \\ %
\textit{MutantX-S} & 4s & 220h & 3h34m  & 41h \\
\hline
\textit{PSS}  & {1h18m} %
& 190h & 2h9m &  {263h} \\ %
\qquad preproc. &  {(1h18m)} %
& (81h) & (16m42s) &  {(233h)} \\ %
\textit{PSS$_O$}  & {15m8s}%
& 116h  & {2h12m}  & 31h29m \\ %
\qquad preproc.  & {(15m6s)}%
& (14h3m)  & {(33m3s)} & (5h23m) \\% IoT: 47m => 33m3s (erreur prepro. de samples non utilises)
\hline
\hline    
\textit{LibDX} & 1m4s  &  1965h  &	7h47m & 170h \\
\hline
\textit{StringSet} & 38s   &  542h  & 9h21m & 253h \\
\textit{FunctionSet}  & 3s & 37h & 7m47s & 27h34m \\
\hline
\end{tabular}
\end{table}

\begin{table}
\centering
\caption{(RQ1) Runtimes per clone search (sec). Include preprocess.~time. Significant preprocess.~times reported in "( \ )".}
\label{runtimesCS}
\begin{tabular}{|l|c|c|c|c|}
\hline
Dataset &  Basic & BinKit & IoT & Windows \\
{\# Programs} &   1K & 98K & 20K & 85K \\
\hline
\textit{B$_{size}$} &  $< 0.01$  & 0.11 & 	0.14 &  0.63\\
\textit{D$_{size}$}  & $< 0.01$  & 0.11 & 	0.14 &  0.63\\
\textit{Shape} & 0.02  &  0.05  &	0.06   & 0.31 \\
\hline
\textit{ASCG} & {\small 1.42 (1.42)}  &  {\small 0.37 (0.21)}   &  {\small 0.25 (0.06)} &  {\small {17.68 (16.60)}} \\
\textit{MutantX-S} & $< 0.01$  & 0.57  &	0.64 & 3.00 \\
\hline
\textit{PSS}  &  {\small 1.41 (1.41)} &  {\small 0.49 (0.21)} &  {\small 0.39 (0.05)}   &  {\small {19.17 (16.95)}} \\
\textit{PSS$_O$}  &  {\small 0.27 (0.27)} &  {\small 0.30 (0.04)}  &   {\small {0.40 (0.10)}} &  {\small 2.29 (0.39)}\\ %
\hline
\hline    
\textit{LibDX} & 0.02  &  5.09 & 1.40 & 12.43 \\
\hline
\textit{StringSet} & 0.01  &  1.40 & 1.69 & 18.47 \\
\textit{FunctionSet}  &   $< 0.01$             & 0.10 &{0.02} & 2.01 \\
\hline
\end{tabular}
\end{table}

\subsection{RQ1: Evaluation of Speed}
\label{RQSpeed}
We report in Table~\ref{runtimesTotal}  the runtimes and the preprocessing time on each dataset to be fully comprehensive. 

\myparagraph{Basic} On the Basic dataset containing  950 programs, 
our method is the slowest and takes {1h18m}.
Nearly everything is spent during the prepossessing.
The adapted spectral method for call graph {\it ASCG} has similar runtimes. \PSSO takes only {15m8s}%
, the optimization dividing the runtime of \PSS by  more than 5.  {\it LibDX} takes 1m4s, and {\it StringSet} 38s. The N-gram method {\it MutantX-S} and the {\it FunctionSet} method are very fast and take less than ten seconds.

\myparagraph{BinKit} On the BinKit dataset, which contains 97,760 programs of an average size of 313 Ko, \PSS takes 190h,  \PSSO  116h, and  {\it MutantX-S} is slower with 220h. 
With literal identifiers, {\it LibDX}   and   {\it StringSet} are much slower (1965h and 542h, resp.). 
 {\it FunctionSet} is fast (37h).

\myparagraph{IoT Malware} On the IoT dataset containing  19,959 IoT malware, \PSS takes only 2h9m. It is faster than {\it MutantX-S} (3h34m). 
Surprisingly, \PSSO is a bit slower than \PSS and takes {2h12m}. Among methods using literal identifiers, {\it FunctionSet} is fast, with less than 8 minutes in total. {\it LibDX} takes 7h47m, while {\it StringSet} is the slowest with 9h21m.

\myparagraph{Windows} {\PSS takes 263h on the Windows dataset. That is far higher than {\it MutantX-S} ($41$h) and a bit higher than {\it StringSet} ($253h$h). However, \PSSO takes less than $32$ hours.}
Table~\ref{runtimesCS} reports average runtimes per clone search. We can see that \PSS preprocessing  time can sometimes be important, e.g., on large Windows binaries ( {16.95s} %
on similarity checks). First, note that  preprocessing time does not increase with the repository size. Second, \PSSO is especially  optimized for such cases, and its preprocessing time remains low in all cases.

\begin{mybox}{Conclusion (RQ1)}
\PSS is often roughly as fast  as {\it MutantX-S} on larger datasets, yet it struggles on  large Windows programs. \PSSO remedies this default and is consistently faster than other approaches, but the baselines and {\it FunctionSet}. 
{Interestingly, {\it StringSet} is slow on large benchmarks.}

\end{mybox}

\begin{table}[htbp]
\caption{(RQ2) Precision scores}
\label{scoresdatasets}
\begin{tabular}{|l|c|c|c|c|}
\hline
{Dataset} &   Basic & BinKit & IoT & Windows \\
\hline
$B_{size}$ &  0.17 & {0.166} & 0.819  & 0.196 \\  %
$D_{size}$ &  0.16 & {0.062} & 0.787 & 0.445 \\ %
{\it Shape} &  0.19 & {0.297} & 0.818 & {0.389} \\ 
\hline
{\it ASCG} &  0.24  & {0.554} & 0.759& 0.444 \\ %
{\it MutantX-S} &  0.38 & {0.354} & 0.870 & 0.472 \\ %
\hline
\PSS  &  0.38 & {0.619} & 0.863 &  0.475 \\ %
\PSSO &  0.38 & {0.619} & 0.862 &  0.466 \\ %
\hline
\hline
{\it LibDX} &  0.70 & {0.882} & 0.707 &  0.044 \\ %
\hline
{\it StringSet} &  0.94 & 0.970 & 0.922  & 0.501 \\ %
{\it FunctionSet} &  0.87 & {0.500} &  {0.644}  &  0.426 \\
\hline
\hline
Random &  0.02 & 0.004 &  0.477  &  $< 0.001$ \\
\hline
\end{tabular}
\end{table}

\subsection{RQ2: Evaluation of Precision}
\label{RQPrecision}
We compute precision scores on each dataset. We report the results in Table~\ref{scoresdatasets}.

\myparagraph{BinKit} \PSS and \PSSO attain a score of {$0.619$} on BinKit, while the other spectral method {\it ASCG} has only {$0.554$}.  {\it MutantX-S} is well behind with {$0.354$}. In fact, we show in Table~\ref{tab:scoreshighlight} that it achieves scores of $0.01$ in cross-architecture scenarios as well as against obfuscations.  With literal identifiers, {\it StringSet} attains $0.970$ and {\it LibDX} {$0.882$}. The {\it FunctionSet} method has only a score of {$0.500$}.

\myparagraph{IoT Malware} \PSS has a score of $0.863$, close to {\it MutantX-S} ($0.870$). \PSSO is very close with $0.862$, while {\it ASCG} attains $0.759$. With literal identifiers, {\it StringSet} achieves a score of $0.922$. Other literal identifier methods have some troubles. {\it FunctionSet} has a score of {$0.644$} because only very few external names are available. Moreover, {\it LibDX} attains $0.707$ because {\it LibDX} extracts constant string values from read-only sections, which are scarce inside IoT firmware.

\myparagraph{Windows} \PSS attains  a score of $0.475$ on Windows, just above {\it MutantX-S} ($0.472$) and well above {\it ASCG} ($0.444$). \PSSO is a bit behind \PSS and {\it MutantX-S} with $0.466$.
Among methods with literal identifiers,  {\it StringSet} attains $0.501$.  {\it LibDX} attains only $0.044$. Again, {\it LibDX} extracts constant string values from well-defined read-only sections, which are not prevalent in Windows programs. As before, {\it FunctionSet} has a rather low score here of only $0.426$.

\begin{mybox}{Conclusion (RQ2)}
\PSS and \PSSO are usually as precise as {\it MutantX-S} except in  cross-architecture and obfuscations scenarios, for which {\it MutantX-S} fails.
When literal identifiers are meaningful, {\it StringSet} is the most precise method in all datasets, while {\it FunctionSet} and {\it LibDX} struggle on IoT and Windows datasets.
 \end{mybox}

 \begin{table*}
    \caption{(RQ2,RQ3) Precision scores on the BinKit dataset} 
    \label{tab:scoreshighlight}
\begin{tabular}{|l|cccccc|ccc|cccc|cccc|}
\hline
Category &
\multicolumn{6}{|c|}{Optimization level} &
\multicolumn{3}{|c|}{Cross-compiler} &
\multicolumn{4}{|c|}{Cross-architecture} &
\multicolumn{4}{|c|}{vs. Obfuscation$\dagger$}
\\
\hline
{}   &  O0 & O0 & O0 & O1 & O1 & O2 & gcc-4 & clang-4 & clang & arm & arm & mips & 32 & & & & \\
{vs.} &   O1 & O2 & O3 & O2 & O3 & O3 & gcc-8 & clang-7 & gcc & mips & x86 & x86 & 64 & bcf & fla & sub & all \\
\hline
$B_{size}$      &  0.04 & 0.04 & 0.07 & 0.19 & 0.11 & 0.21 & 0.11 & 0.45 & 0.07 & 0.03 & 0.10 & 0.04 & 0.04 & 0.04 & 0.01 & 0.08 & 0.01 \\
$D_{size}$      & 0.03 & 0.03 & 0.03 & 0.06 & 0.05 & 0.07 & 0.07 & 0.09 & 0.04 & 0.02 & 0.05 & 0.03 & 0.04 & 0.02 & 0.01 & 0.05 & 0.01\\
{\it Shape}     & 0.19 & 0.07 & 0.06 & 0.17 & 0.11 & 0.33 & 0.38 & 0.65 & 0.16 & 0.04 & 0.16 & 0.04 & 0.19 & 0.25 & 0.27 & 0.48 & 0.23 \\
{\it ASCG}      & 0.40 & 0.12 & 0.10 & 0.43 & 0.24 & 0.68 & 0.78 & 0.91 & 0.46 & 0.08 & 0.46 & 0.06 & 0.59 & 0.54 & 0.64 & 0.78 & 0.48 \\
{\it MutantX-S} &  0.04 & 0.03 & 0.03 & 0.43 & 0.36 & 0.64 & 0.67 & 0.80 & 0.14 & 0.02 & 0.01 & 0.01 & 0.06 & 0.09 & 0.03 & 0.54 & 0.01 \\
\hline
\PSS            & 0.54 & 0.23 & 0.17 & 0.59 & 0.38 & 0.70 & 0.79 & 0.91 & 0.51 & 0.39 & 0.55 & 0.39 & 0.66 & 0.53 & 0.57 & 0.82 & 0.46 \\
\PSS$_O$        & 0.53 & 0.24 & 0.17 & 0.60 & 0.39 & 0.68 & 0.78 & 0.90 & 0.51 & 0.44 & 0.54 & 0.44 & 0.66 & 0.52 & 0.56 & 0.82 & 0.46 \\
\hline
\hline
{\it LibDX}      & 0.89 & 0.89 & 0.89 & 0.89 & 0.89 & 0.89 & 0.89 & 0.86 & 0.78 & 0.87 & 0.89 & 0.90 & 0.88 & 0.87 & 0.86 & 0.86 & 0.86 \\
\hline
{\it StringSet} & 0.97 & 0.97 & 0.97 & 0.97 & 0.97 & 0.97 & 0.97 & 0.97 & 0.97 & 0.96 & 0.98 & 0.96 & 0.97 & 0.96 & 0.97 & 0.96 & 0.97 \\
{\it FunctionSet}&  0.55 & 0.53 & 0.53 & 0.55 & 0.55 & 0.56 & 0.46 & 0.68 & 0.55 & 0.29 & 0.02 & 0.00 & 0.23 & 0.61 & 0.61 & 0.61 & 0.61 \\
\hline
\end{tabular}
    \medskip

Random clone search results in a precision score inferior to $0.005$ on all test fields.

$\dagger$: The  BinKit dataset does not consider any obfuscation of literal identifiers

\end{table*}

\begin{table}[htbp] %
\centering
\caption{(RQ3) Average rank-biserial correlation for $H$}
\label{tab:correlations}
\begin{minipage}{.5\linewidth}
\centering
\begin{tabular}{|l|c|}
\hline
 Framework  & Basic dataset  \\
\hline
$B_{size}$  & \textbf{0.02} \\
$D_{size}$  & \textbf{0.01} \\
{\it Shape} &  \textbf{0.04} \\
\hline
{\it ASCG}  &   \textbf{0.08} \\
{\it MutantX-S} & {0.33} \\
\hline
\end{tabular}
\end{minipage}%
\begin{minipage}{.5\linewidth}
\centering
\begin{tabular}{|l|c|}
\hline
 Framework  & Basic dataset \\
\hline
\PSS     &   \textbf{0.06} \\
\PSS$_O$ &  \textbf{0.06} \\ 
\hline
\hline
{\it LibDX} &  \textbf{-0.07} \\
\hline
{\it StringSet}  &          {0.45} \\
{\it FunctionSet} &          {0.20} \\
\hline
\end{tabular}
\end{minipage} 

\end{table}

\subsection{RQ3: Evaluation of Robustness}
\label{RQRobustness}
The last evaluation measures the robustness of the ten clone search methods that survived the speedtest. For this, we consider four scenarios with (i) cross-optimization, (ii)  cross-compiler, (iii) cross-architecture and (iv) in the presence of obfuscations. The evaluation leans on the BinKit dataset that we presented earlier.

\myparagraph{Results} We report the most crucial test field scores in Table~\ref{tab:scoreshighlight}. 
When literal identifiers are available, {\it StringSet} and {\it LibDX} are very stable in all scenarios. {\it FunctionSet} is stable except in scenarios involving cross-architecture because  external function names differ between architectures. Note that a strong limitation to this finding is that the considered obfuscations do not hide nor encrypt literal strings and external calls (API), while it is common practice.

\PSS and \PSSO are much more robust than {\it MutantX-S} in cross-architecture, cross-optimization and obfuscations scenarios. For instance, {\it MutantX-S} falls to $0.02$ from the arm to mips architecture, while \PSS maintains a score of $0.39$. The more basic spectral method {\it ASCG} also falls to $0.08$ in this scenario.
Interestingly, \PSS and \PSSO perform better in the cross-architecture test fields than in the  (-O0, -O3) and (-O0, -O2) test fields. We hypothesize that while the architecture does not impact that much  the produced call graph, advanced optimizations do --  function inlining  is precisely turned on by the -O2 optimization level in both GCC and Clang.

\myparagraph{Statistical analysis} A common pitfall of similarity detection is that a method could in the end consider as  similar two programs based on  some side aspects (e.g., architecture, compiler used or optimization version) irrelevant from the clone search point of view.  We  evaluate the sensitivity of the different approaches to such bias by computing rank-biserial correlations between (a) similarity rank in new clone searches and (b) sharing an optimization level.
We report average correlations in Table~\ref{tab:correlations} (the lower, the better and less sensitive).   \PSS, \PSSO, {\it ASCG} and {\it LibDX} have very small correlations of less than $0.10$. On the other hand, the {\it StringSet} method correlation is moderate ($0.45$), indicating some bias.
Surprisingly, this bias does not seem to impact the robustness of {\it StringSet}  (Table~\ref{tab:scoreshighlight}).
The N-gram method {\it MutantX-S} has a lower correlation  of $0.33$ and {\it FunctionSet} has a small correlation of $0.20$. 
\begin{mybox}{Conclusion (RQ3)}
\PSS and \PSSO are robust to cross-optimization, cross-compiler, cross-architecture and obfuscations scenarios, while  
{\it MutantX-S} suffers significant precision loss in the  cross-architecture and obfuscations cases.  
\end{mybox}

\begin{table}[htbp] %
\caption{(RQ4) Components precision scores}
\label{tab:ablation:scores}
\begin{tabular}{|l|cc|c|c|}
\hline
{Dataset} &   Basic & BinKit & IoT & Windows \\
\hline
{\it simCG}  &   0.29  & 0.596 & 0.856  & 0.459   \\ 
{\it simCFG}  &   0.29  & 0.424 & 0.856  & 0.163  \\ 
\hline
\PSS  &  0.38 & {0.619} & 0.863 &  0.475 \\ %
\hline
\end{tabular}
\end{table}

\begin{table}[htbp]
\centering
\caption{
(RQ4) Components runtimes per clone search (sec).
{\footnotesize Include preprocess.~time. Significant preprocess.~times reported in "( \ )".}
}
\label{tab:ablation:runtimes}
\begin{tabular}{|l|c|c|c|c|}
\hline
Dataset &  Basic & BinKit & IoT & Windows \\
\hline
\textit{simCG}  & 1.41 (1.41) & 0.36 (0.21) & 0.22 (0.05)   &  {18.07 (16.95)} \\
\textit{simCFG}  & $< 0.01$ &  0.14 &  0.16  &  1.06 \\
\hline
\textit{PSS}  & 1.41 (1.41) & 0.49 (0.21) & 0.39 (0.05)   & {19.17 (16.95)} \\
\hline
\end{tabular}
\end{table}

\subsection{RQ4: Ablation Study}
In Table~\ref{tab:ablation:scores}, we report the precision scores of the two components of PSS: {\it simCG} and {\it simCFG}. The first is a comparison between eigenvalues of the call graph, while the second is a comparison between the number of edges of functions control flow graphs. \PSS always attains a higher precision score than {\it simCG} and {\it simCFG} on every dataset. We remark that {\it simCFG} alone is not precise on the Windows dataset ($0.163$ vs. $0.459$ for {\it simCG}).
In Table~\ref{tab:ablation:runtimes}, we report each component's average runtimes per clone search. As expected, \PSS runtimes are the addition of {\it simCG} and {\it simCFG} runtimes. Therefore, \PSS is at worse one second slower than {\it simCG}.

 \begin{mybox}{Conclusion (RQ4)}
\PSS is more precise than its components for the price of a slight increase in runtimes.
 \end{mybox}

\begin{table}[htbp]
\caption{Informal summarized comparison \label{tab:comparison}}
\begin{tabular}{|c|c|c|c|c|}

\hline
Method        &   speed     & precision & robust.  & beware \\ 
\hline
{\it ASCG}~\cite{GSA}           &   +         &    -       &     +      &   \\  
{\it MutantX-S}~\cite{MUTANTXS}     &     +        &    +       &     {-}-      &   \\
\hline
\PSS/\PSSO     &   +/++         &    +       &     +      &   \\  
\hline
\hline
{\it LibDX}~\cite{LibDX}        &    -         &    ++     &    ++        & str. extraction  \\ 
             &              &          &            & str. obf.  \\ 
\hline
{\it StringSet}    &    {-}-         &    +++     &    ++        & str. obf.  \\ 
{\it FunctionSet}  &    +++        &    -       &    -        & fun. name  obf. \\
              &            &          &            &   static linking\\
\hline
\end{tabular}
\end{table}

\subsection{Summary of Our Main Results}

  Our novel  spectral methods \PSS and \PSSO reach a sweet spot regarding the trade-off between speed, precision  and robustness.  They {do} not need any training phase,  scale very well to large repositories and are very robust, even in cross-architecture or cross-compiler scenarios and in case of lightweight obfuscation.   Therefore, they are the best candidates  for intensive program clone search. Also, it is worth mentioning that direct adaptations of graph based spectral methods lack precision compared to \PSS, and that  the optimization  \PSSO is necessary over large programs.  
A summarized informal comparison with other methods is given in Table 
\ref{tab:comparison}.

This large study also allowed us to highlight that most prior approaches in the field~\cite{Asm2Vec,GEMINI,SAFE,AlphaDiff}, mostly focused on \textit{function-level} similarity, are far too slow for  \textit{program} clone search.

\section{Related works}
Binary code similarities are extensively studied. As a testimony, the review of Haq and Caballero~\cite{SBC} reports numerous input and output granularities on which to study similarities. 

\myparagraph{Pioneering approaches}
Dullien in 2004~\cite{SCEO} introduced a graph based program diffing approach that constructs  a call graph isomorphism. A follow-up~\cite{GBCEO} extended it to match basic blocks inside matched functions. These two results are the basis for the popular BinDiff program diffing plugin for the IDA disassembler. BinDiff aims to recognize similar binary functions among two related executables. 
In 2006, Kruegel et al.~\cite{PWDSI} presented an approach based on coloring small graphs with fixed size from the control flow graph to identify structural similarities between different worm mutations. 
In 2008, Gao et al. proposed BinHunt~\cite{BINHUNT}  to find differences between two versions of the same program.  BinHunt employs symbolic execution with a constraint solver to prove that two basic blocks implement the same functionality. %

\myparagraph{Program similarity}
The few recent works about program-level similarity~\cite{SPAIN,MSBB}  have already been thoroughly discussed.  
Still, we can mention a few more approaches. 
N-gram methods compare instruction sequences~\cite{EXPOSE,MUTANTXS,MBC,IDEA}. While we could have employed more fine-grained methods than {\it MutantX-S}~\cite{MUTANTXS} -- for example Expos\'e~\cite{EXPOSE} considers trigrams inside a function matching, it quickly leads to serious scalability issues.  
Some other works explore similarities based on dynamic executions and input-output observations~\cite{GBMDDA,ILINE,BEAGLE,BinSim}. Nevertheless, it is hard to thoroughly explore the execution space with dynamic traces -- leading to poor precision, and handling large code repositories requires automating the task of  detecting the sources of input and output of all programs in the repository, which can be very complicated.  
Bruschi et al.~\cite{DETECTSELFMUTATING} tackle the problem of detecting some malware inside a program by matching control flow graphs. But, again, this approach suffers from scalability issues (in the size of the programs) and is thus not amenable to the search over large code repositories.  
The symbolic method by Luo et al.~\cite{SYMBOLICTHEOREMPROVING} is robust to simple obfuscations as well as simple changes. However, the running time of symbolic execution is a critical issue on large programs, and anti-analysis obfuscation hinders symbolic approaches~\cite{ObfuscationSymobolic,KillSymbolic}.  
We have already studied the matching method CGC~\cite{SMM}.
The complex matching by Xu et al.~\cite{SMM}  outperforms a baseline based on external function names and mnemonics. However, we propose {\it StringSet}, a faster, highly precise method comparing sets of constant string values.
A few other matching approaches~\cite{MAV,MatchingFunctions,BinSlayer} share the same strengths and weaknesses.

\myparagraph{Function similarity}
The last five years have seen a tremendous increase in the popularity of binary function similarity with machine learning~\cite{Asm2Vec,AlphaDiff,CODEE,GEMINI,SAFE}. 
Yet,  as already discussed, these methods lead to poor scalability when applied to a program similarity setting. 
More expensive  methods than function embeddings do exist. Notably, dynamic analysis seeks to build upon the semantics of binary codes instead of their mere structural properties. 
BinGo~\cite{BINGO} analyzes various execution traces with concepts such as pruning. 
The work of Hu et al.~\cite{BinaryCodeCloneDetection} emulates binary functions to create semantic signatures. 
Pewny et al.~\cite{CrossBugSearch} propose to translate binary code to an intermediate representation. This representation allows observing inputs and outputs of basic blocs.
These frameworks suffer from the already mentioned pitfalls of dynamic execution: the exploration is either imprecise or very slow. Furthermore, it is unclear how to lift these methods to the case of program similarity, as comparing all functions between multiple codes is costly. 
Built on the idea of intermediate representation, several  approaches~\cite{Tracelet,LeveragingSemantic,Juice} perform simplification before comparing. In FirmUp~\cite{FirmUp}, the matching between intermediate representations incorporates multiple functions.  The formula has to be transformed into an embedding. The larger the code segment it represents, the less precise the embedding is.
Finally, other feature selection methods have been investigated: Rendez-vous~\cite{RendezVous} extracts statistical features at various granularities,  while discovRE~\cite{DISCOVRE} and Genius~\cite{SGBBS} extract features such as the number of arithmetic instructions. Gemini~\cite{GEMINI}  leverages static features from Genius into a machine learning framework.

\myparagraph{Source code similarity}
Computing source similarities can be performed with different structures such as Abstract Syntax Trees~\cite{ASTP,ASTBIEN,SourceCodeDetectionComparison} or Program Dependency Graphs~\cite{SourceCodeDetectionComparison}. It is also possible to normalize instructions and compare code fragments~\cite{ReDeBug,SourceCodeDetectionComparison}. Matching tokens, fragments and structures is effective because there is no compiler optimization step which would introduce variations. Moreover, critical information such as types are lost by compilation, while data dependencies are harder to retrieve on binary programs.

\myparagraph{Graph similarity}
A key question in program similarity is how to compare graphs efficiently. 
New suggestions for graph similarities include novel graph kernels~\cite{DFGS,SKGCA,OVOAK} and the use of machine learning to approximate intractable properties such as graph edit distance~\cite{SIMGNN, GMN, LGEDGNN}. 
Recently, the work of Bay-Ahmed et al.~\cite{JSSMGC} introduced a new graph similarity metric incorporating both spectral information from the Adjacency Matrix and from the Laplacian. 
Moreover, the work of Crawford et al.~\cite{GraphStructureSimilarity}  proposed spectral analysis as a similarity metric of real-world networks.
 Furthermore, the study of Fyrbiak et al.~\cite{GSA} reveals that spectral analysis can compete with more energy-intensive approaches such as GED.

\myparagraph{Library identification}
The pioneering BAT~\cite{BAT} has proposed three methods for library identification, based on strings, compression algorithms and edit distances between bit sequences. They report that edit distance computations are too costly, while strings can be easily obfuscated. OSSPolice~\cite{OSSPolice} has developed similarity measures based on strings.
The special structure of Java programs allows the use of properties such as class and package inclusions~\cite{AndroidLibrary, AndroidBloom} in order to identify Android libraries.

\section{Discussion and Limitations}

While \PSS and \PSSO perform well in our experiments, there are still a number of potential corner cases that must be considered. 
Generally speaking, these methods will suffer on program clones with very different call graphs. Such differences could come for example:  
(1) from significant source code revisions -- it is why we  support only incremental changes of an application or library, 
(2) or from aggressive inter-procedural compiler optimizations, such as function inlining or function sharing -- link-time optimizations may be a growing problem here, 
(3) or from aggressive inter-procedural obfuscation schemes, such as function merging  or virtualization.

Also, as the programs we consider mainly come from  C/C++ source codes, it would be interesting to evaluate all the considered methods over programs written in emerging programming languages (e.g., Rust, Go) that may contain language-specific function call patterns.

\section{Conclusion}

We consider the problem of searching program clones in large code repositories. 
While most prior works have been devoted to function clones, the few existing techniques for program similarity suffer either from scalability issues, low precision,  or low robustness to code variations. 
We propose a novel method called Program Spectral Similarity (\PSS, and especially its optimized version \PSSO) that reaches  a sweet spot in terms of speed, precision, and robustness -- even in cross-compiler or cross-architecture setups.

\begin{acks}
This work is partially supported by: 
a French PIA grant "Lorraine Université d'Excellence"  ANR-15-IDEX-04-LUE, 
 an EU Horizon 2020 research and innovation grant  No 830927 (Concordia),
and  an ANR grant under France 2030 ``ANR-22-PECY-0007’'.

Experiments presented in this paper were carried out using the Grid'5000 testbed, supported by a scientific interest group hosted by Inria and including CNRS, RENATER and several Universities as well as other organizations (see \url{https://www.grid5000.fr}).

\end{acks}

\newpage
\bibliographystyle{ACM-Reference-Format}
\balance
\bibliography{rb}

\end{document}